\documentclass[%
 reprint,
superscriptaddress,
 amsmath,
 amssymb, 
 aps,  
]{revtex4-2}

\usepackage{graphicx}
\graphicspath{{plots/fig1/}{plots/fig2/}{plots/fig3/}{PDF_output/}{Final_plots/}}
\usepackage{comment}
\usepackage{dcolumn}
\usepackage{bm}
\usepackage{hyperref}

\usepackage{xfrac}
\usepackage{siunitx}
\usepackage{orcidlink}
\usepackage[T1]{fontenc} 
\usepackage[dvipsnames]{xcolor}
\usepackage{ulem}
\usepackage{tikz}
\usepackage{upgreek}

\colorlet{ForestGreen}{Black}
\begin{document}

\preprint{APS/123-QED}
\title{Waveguide-integrated colour centres in silicon carbide with broadband photonic crystal reflectors for efficient readout}

\author{Marcel Krumrein\,\orcidlink{0009-0009-4685-8831}}
 \affiliation{3rd Institute of Physics, IQST, and Research Center SCoPE, University of Stuttgart, 70569 Stuttgart, Germany}

\author{Julian M. Bopp\,\orcidlink{0000-0003-0370-9114}}
 \affiliation{Humboldt-Universität zu Berlin, Department of Physics, 12489 Berlin, Germany}
 \affiliation{Ferdinand-Braun-Institut gGmbH, Leibniz-Institut für Höchstfrequenztechnik, 12489 Berlin, Germany}

\author{Timo Steidl\,\orcidlink{0009-0000-2819-3228}}
 \affiliation{3rd Institute of Physics, IQST, and Research Center SCoPE, University of Stuttgart, 70569 Stuttgart, Germany}
 
\author{Wolfgang Knolle}
 \affiliation{Leibniz-Institute of Surface Engineering (IOM), 04318 Leipzig, Germany}

\author{Jawad Ul-Hassan}
 \affiliation{Department of Physics, Chemistry and Biology, Linköping University, SE-581 83 Linköping, Sweden}

\author{Vadim Vorobyov\,\orcidlink{0000-0002-6784-4932}}
\email[]{v.vorobyov@pi3.uni-stuttgart.de}
 \affiliation{3rd Institute of Physics, IQST, and Research Center SCoPE, University of Stuttgart, 70569 Stuttgart, Germany}
 
\author{Tim Schr\"oder}
 \affiliation{Humboldt-Universität zu Berlin, Department of Physics, 12489 Berlin, Germany}
 \affiliation{Ferdinand-Braun-Institut gGmbH, Leibniz-Institut für Höchstfrequenztechnik, 12489 Berlin, Germany}

\author{J\"org Wrachtrup\,\orcidlink{0000-0003-3328-9093}}
 \affiliation{3rd Institute of Physics, IQST, and Research Center SCoPE, University of Stuttgart, 70569 Stuttgart, Germany}
 \affiliation{Max Planck Insitute for Solid State Research, 70569 Stuttgart, Germany}


\begin{abstract}
Spin-active colour centres in 4H silicon carbide are promising candidates as building blocks for quantum information applications. 
To increase the photon count rate of the emitters at low temperatures, the colour centres must be integrated into nanophotonic structures and characterised under cryogenic conditions. 
Here, we design and fabricate waveguide structures attached with an efficient Dinosaur photonic crystal reflector at one side.
The devices show broadband reflection over a range of $60\,$THz with a peak reflectance above $80\%$.
Additionally, colour centres were integrated into these structures and characterised at cryogenic conditions.
The emission was collected by a tapered-waveguide-tapered-fibre interface.
Although the spectral stability of the emitters must be further improved for high excitation powers, the saturation intensity in standard PLE measurements is $I_\text{s,LT}=(103.8\pm4.2)\,\text{kcps}$.
The count rate can be further improved to about $125\,$kcps with a charge-resonance check measurement scheme.
To highlight the relevance of our devices, we theoretically show that these count rates enable optical single-shot readout with a fidelity exceeding $98\%$.
\end{abstract}

\maketitle

\section{Introduction}

Silicon carbide (SiC) is an emerging complementary metal-oxide-semiconductor (CMOS) compatible material system hosting promising quantum emitters for applications in quantum technologies \cite{Castelletto.2020,Castelletto.2022}.
In particular, the silicon vacancies V1 (h-V$_\text{Si}$) and V2 (k-V$_\text{Si}$) gained attraction due to their excellent spin-optical properties.
Both colour centres show long spin-coherence times \cite{Widmann.2015,Simin.2017} as well as lifetime-limited optical linewidths for temperatures up to $20\,$K \cite{Udvarhelyi.2020}.
The coupling to nearby nuclear spins allows for nuclear-spin-assisted single shot readout (SSR) of the electron spin \cite{Lai.2024,Hesselmeier.2024}, an important step towards efficient quantum protocols.
However, the direct spin readout via the optical transition is still a challenge. 
Therefore, it is important to increase the photon detection efficiency of the V2 colour centre.
For this, the integration into photonic structures like solid immersion lenses \cite{Sardi.2020,Bekker.2023}, pillars \cite{Radulaski.2017}, antennas \cite{Korber.2024}, and a Fabry-Perot microcavity \cite{Hessenauer.2025} was shown.
In silicon carbide, none of these systems achieve sufficient photon collection efficiency; consequently, single-shot readout of the electron spin has not been demonstrated yet.
A promising approach is to integrate the emitters into waveguide structures, where photon count rates of about $180\,$kcps were achieved using a tapered-waveguide-tapered-fibre (TWTF) interface for photon collection \cite{Krumrein.2024,Burek.2017}.
At ambient conditions, the waveguide can be approached with optical fibres from both ends to obtain maximum collection efficiency.
However, due to limited space in a common cryostat the access with two facing fibres is not feasible and, consequently, half of the photons is typically lost.
Additionally, combining the photons from two single-mode fibres efficiently without introducing a time delay in any of the paths would be challenging.
By adding a photonic crystal (PhC) reflector to the waveguide's open end, the photon count rate can be ideally doubled.
To increase the collection rate of both the colour centre's narrow zero-phonon line (ZPL) and spectrally broad phonon sideband (PSB) emission, the PhC reflector has to maintain a high reflectance across a broad spectral operating range.
For SiC nanostructures with a triangular cross section, a hole-based reflector has been proposed recently \citep{Majety.2023}.
We show that corrugation-based PhC nanostructures with a triangular cross section, which are called Dinosaur nanostructures \cite{Bopp.2025}, provide the same peak reflectance but extend the spectral operating range to about $60\,$THz if a tapered waveguide-reflector interface is introduced -- comparable to the tapered waveguide-cavity interfaces of Sawfish PhC cavities with corrugation features and a rectangular cross section \citep{Bopp.2024, Pregnolato.2024}.
Realizing such a tapered interface is possibly more challenging for hole-based PhC reflectors due to the required small hole diameters \citep{Palamaru.2001}.

\begin{figure*}
\includegraphics[width=0.95\textwidth]{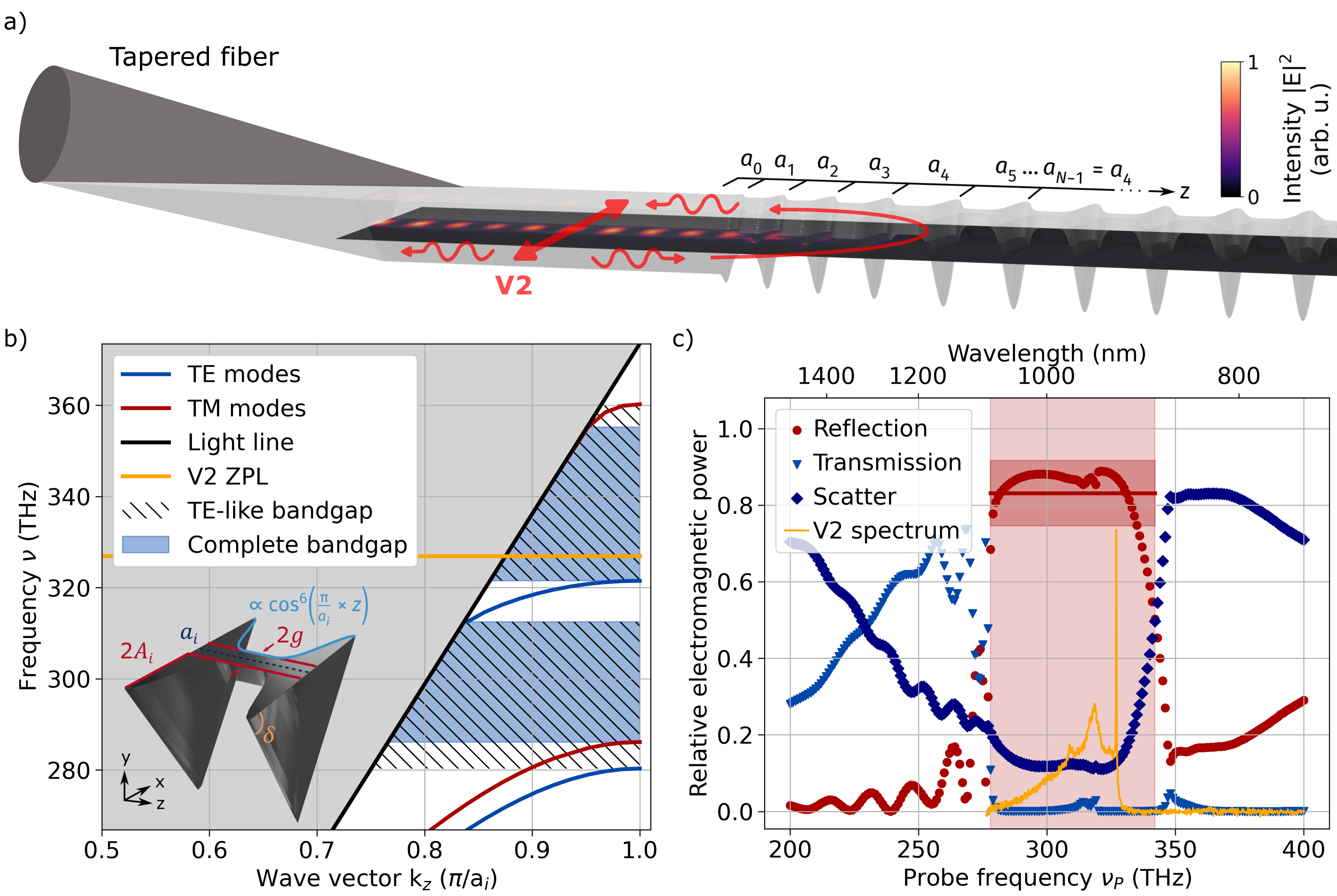}
\caption{Illustration and simulations of the Dinosaur reflector.
(a) PhC reflector with attached waveguide section and a tapered-waveguide-tapered-fibre (TWTF) interface to collect the emitted photons of an integrated V2 colour centre.
The interface between the triangular waveguide and the attached Dinosaur reflector comprises a tapered section, involving $i\in\left[0,5\right)$ unit cells with increasing lengths $a_i$ and corrugation amplitudes $A_i$.
The waveguide-interfacing half of the first unit cell $i=0$ matches the waveguide width.
The electric field intensity of the interference of an incident waveguide mode and its part reflected off the Dinosaur reflector is displayed in the xz-cross section below the reflector surface at a depth that matches one-third of the triangular cross section's height in y-direction.
(b) Stacking identical Dinosaur unit cells (see inset) along the z-direction (for geometry parameters see main text) yields an optical band structure with a set of TE-like (blue lines) and TM-like (red lines) Bloch bands separated by TE-like (dashed areas) and complete (blue-shaded areas) bandgaps within a regime where the wave vector $k_\mathrm{z}$ exceeds the light line $\nu = c|\vec{k}|$ (black line).
Here, $c$ is the speed of light.
The ZPL of a V2 colour centre (yellow line) is reflected by the upper complete bandgap.
Inset: Dinosaur unit cell defined by a triangular cross section and a corrugation profile proportional to a cosine function exponentiated by an even integer $e\geq 2$.
(c) The reflection (red dots), transmission (light blue triangles), and scatter (dark blue diamonds) spectra of a tapered Dinosaur reflector consisting of 18 unit cells are obtained by scanning the frequency $\nu_\mathrm{P}$ of an incident probe field in FEM simulations.
The reflector's spectral operating range (shaded in light red) is defined by a reflectance larger than $50\%$.
Within this range, the red line indicates the mean reflectance and the area shaded in dark red represents the corresponding one standard deviation uncertainty.
The orange line shows the V2 emission spectrum.}
\label{fig:figure1}
\end{figure*}

In this work, we simulate and fabricate broadband PhC reflectors which we further integrate into a closed-cycle cryostat and investigate the spectral stability and collection rate of embedded V2 colour centres via a TWTF interface (see Figure \ref{fig:figure1}a).
In photoluminescence excitation (PLE) measurements, the waveguide-integrated emitters show a spectral stability similar to bulk emitters.
Photon count rates exceeding $100\,$kcps are possible in standard PLE measurements which is an increase of one order of magnitude compared to bulk emitters.
Finally, we show that our device can achieve optical single-shot readout of the electron spin.

\section{Results and Discussion}

\subsection{Simulations}

The reflectance of a periodic PhC structure composed of identical unit cells stacked along the z-direction is determined by the band structure of its optical Bloch modes.
The inset in Figure \ref{fig:figure1}b depicts the Dinosaur reflector's unit cell with its geometry parameters such as the length of the $i$-th unit cell $a_i$, the respective corrugation amplitude $A_i$, the corrugation profile exponent $e$, the gap width $g$, and the sidewall angle $\delta$.
Hence, the corrugation profile is $\pm x(z) = 2 A_i \cos^e\left(\mathrm{\pi}/a_i \times z\right)+g$.
Finite element method (FEM) eigenmode simulations performed with the software package \textit{JCMsuite} \citep{JCMsuite} reveal the Dinosaur reflector unit cell's band structure for a parameter set $a_4=401.3\,$nm, $A_4=171.3\,$nm, $e=6$, $g=60.6\,$nm, and $\delta=54^\circ$, as shown in Figure \ref{fig:figure1}b.
Optical modes with wave vectors $k_\mathrm{z}$ parallel to the stacking direction that exceed the light line cannot escape the nanostructure in x- or y-directions.
Instead, they are guided along the nanostructure as Bloch modes.
Fundamental TE- and TM-like bands correspond to dielectric Bloch modes where the electric field is localised to the interfaces between adjacent unit cells, exhibiting elliptical mode profiles with a semimajor axis aligned with the x- and y-axis, respectively.
Higher-order modes possess more complex mode profiles but also guide the light predominantly inside the SiC material, which is characteristic for their dielectric nature.
TE- and TM-like bandgaps separate bands of the respective polarisation and prohibit the transmission of light with the corresponding polarisation and frequencies through the PhC structure.
Complete bandgaps, arising from the overlap of TE- and TM-like bandgaps, are desirable since they enable polarisation-independent reflectors.
For the given parameters, a first TE-like bandgap occurs in the range from $280.3\,$THz to $312.5\,$THz ($10.9\,$\% gap-midgap ratio).
Embedded colour centres with dipoles oriented parallel to the x-axis (which is equivalent to the crystal c-axis) have best mode overlap with the fundamental TE-like band.
Thus, the lower edge of the first TE-like bandgap constitutes a lower limit of the reflector's operating range.
Moreover, there is a lower complete bandgap reaching from $286.1\,$THz to $312.5\,$THz ($8.8\,$\% gap-midgap ratio) and an upper one in the range of $321.5\,$THz to $355.3\,$THz ($10.0\,$\% gap-midgap ratio).
While the lower complete bandgap is suitable for reflecting the V2's PSB emission, its ZPL emission falls into the upper complete bandgap.

Turning the sequence of stacked Dinosaur unit cells into an efficient waveguide-attached Dinosaur reflector demands a tapered interface between the waveguide and the periodic PhC structure to adiabatically interconvert waveguide and Bloch modes \citep{Sauvan.2005}.
Such a tapered interface defined by corrugations that adiabatically build up in conjunction with an increasing unit cell length $a_i$ (Figure \ref{fig:figure1}a) reduces scattering losses and therefore increases the reflectance and spectral operating range.
Considering the tapered reflector's top surface, the waveguide interfaces the reflector at $z=0\,$nm.
For $z<0\,$nm, the waveguide's half width is $303.2\,$nm.
The reflector possesses a yz-symmetry plane at $x=0\,$nm, rendering it sufficient to discuss its upper corrugation profile ($x>0\,$nm).
Hence, the tapered waveguide-reflector interface consists of five unit cells with increasing lengths $[a_0, a_1, a_2, a_3, a_4] = [108.4, 247.2, 299.2, 326.7, 401.3]\,$nm, corrugation maxima of $[x_i^+]=[339.1, 359.2, 371.4, 383.5, 403.2]\,$nm at the interfaces of adjacent unit cells, and corrugation minima of $[x_i^-]=[303.2, 216.1, 167.3, 137.7, 108.5]\,$nm at the unit cell centres connected by $\cos^e$ segments.
The parameter values originate from Bayesian optimisation of the mean reflectance across a frequency range from $290\,$THz to $330\,$THz, using JCMsuite's optimisation toolkit \citep{JCMsuite}.

To assess the reflectance of the full reflector system involving a $5\,\mathrm{\mu}$m long waveguide, the waveguide modes are determined first for each probe frequency $\nu_\mathrm{P}$.
Second, in one scattering simulation for each $\nu_\mathrm{P}$, the respective elliptical waveguide mode that resembles the Dinosaur unit cell's fundamental TE-like Bloch mode is launched into the waveguide along the $+$z-direction.
For a frequency $\nu_\mathrm{P}$ within the spectral operating range, Figure \ref{fig:figure1}a visualises the resulting electric field intensity cross section at a depth below the top surface where colour centres are supposed to be positioned.
Intensity maxima and minima in the waveguide indicate interference between the incident mode and its reflected part.
Notably, the incident field hardly penetrates the reflector.
Reflection, transmission, and scatter spectra are obtained by integrating the resulting electromagnetic energy flux across the computational domain's $-$z-boundary, $+$z-boundary, and the remaining $\pm$x- and $\pm$y-boundaries, respectively.
Finally, the spectra are normalised by the total injected power that is the integral of the electromagnetic energy flux over all computational domain boundary surfaces.
Figure \ref{fig:figure1}c displays the reflection, transmission, and scatter spectra of the optimised waveguide-reflector system composed of 18 unit cells, including the tapered interface.
In agreement with the exponential decay of electromagnetic waves penetrating the band gaps of PhCs \citep{Lin.1993}, i.e., tunneling barriers, we find that the reflector properties have already converged for lengths exceeding 10 unit cells.
Its reflectance rises rapidly around $278(1)\,$THz, where the transmittance through the reflector steeply drops to zero, indicating the lower bound of the spectral operating range that is determined by the lower edge of the first TE-like bandgap (Figure \ref{fig:figure1}b).
The spectral operating range, as defined by a reflectance larger than $50\,$\%, spans a $64(2)\,$THz window which corresponds to a wavelength range of about $200\,$nm (area shaded in light red in Figure \ref{fig:figure1}c).
Within this bandwidth, a mean theoretical reflectance of $83(9)\,$\% is reached (red line).
The reflectance uncertainty is the standard deviation of the reflectance values within the operating range (area shaded in dark red), while the uncertainty of the operating range itself originates from the $1\,$THz sampling rate of the probe frequencies $\nu_\mathrm{P}$.
At $342(1)\,$THz, the reflectance again drops below $50\,$\%, which cannot directly be explained by a band edge in Figure \ref{fig:figure1}b.
At frequencies beyond the spectral operating range, the transmission remains close to zero whereas the scattering increases indicating high optical losses.
We attribute this behaviour to originate from scattering at the tapered waveguide-reflector interface.
Thus, the first four unit cells define the upper bound of the spectral operating range.
Small peaks in the transmission spectrum at $317\,$THz and $348\,$THz may be signatures of higher-order Bloch modes.

\begin{figure*}
\centering
\includegraphics[width=0.95\textwidth]{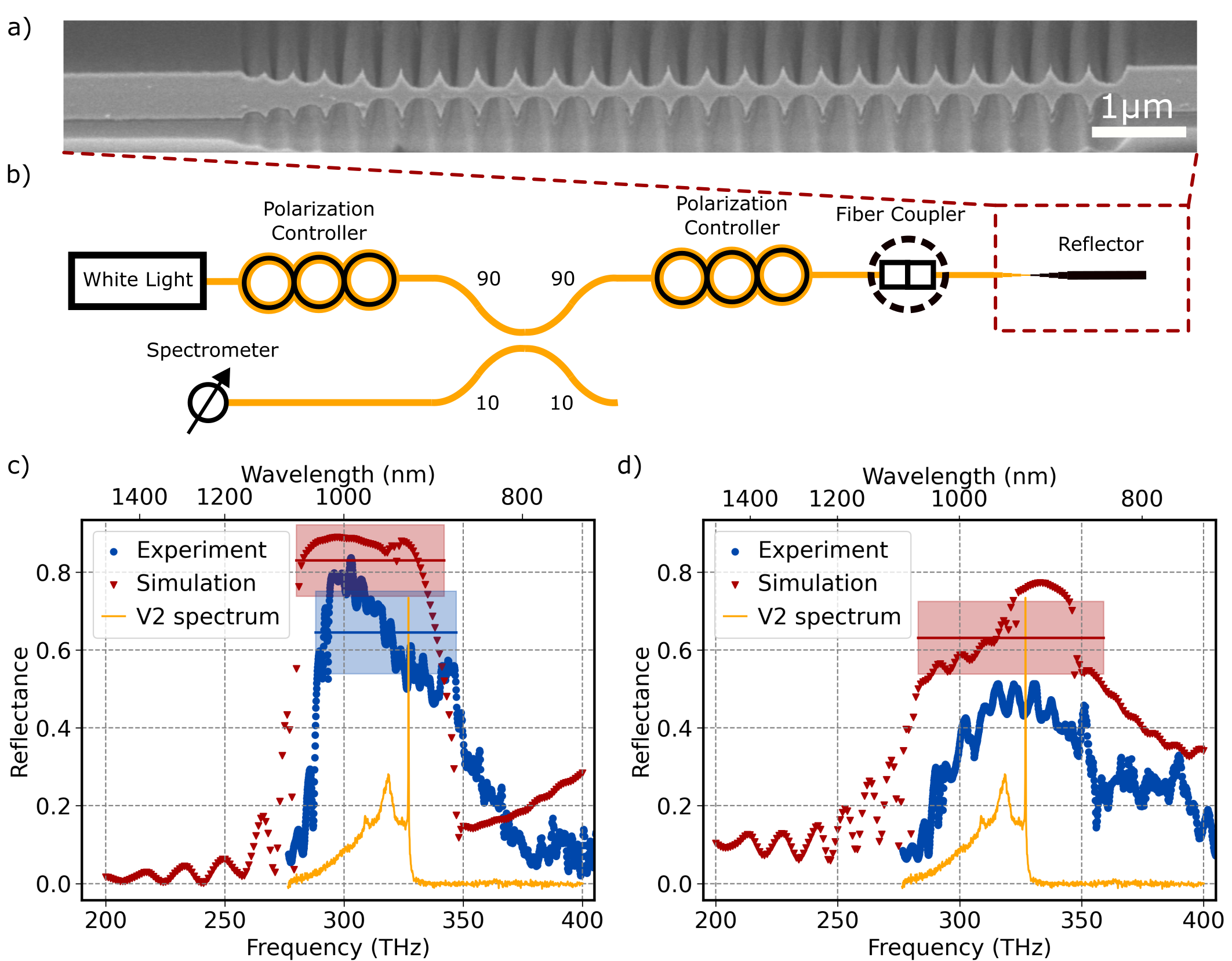}
\caption{Characterisation of fabricated reflector structures. 
(a) SEM image of a fabricated reflector. On the left side, the tapering region is visible.
(b) Setup to measure the reflectance of the devices. White light passes a 90:10 beamsplitter and is sent to a tapered optical fibre which is attached to the taper of the reflector. The reflected photons are measured by a spectrometer in the second arm of the beamsplitter. Polarisation controllers are included because the beamspliter and the reflector are polarisation-dependent. 
(c,d) Measured (blue dots) and simulated (red triangles) reflection spectra of a fabricated waveguide-attached reflector (c) with and (d) without a tapered waveguide-reflector interface.
Areas shaded in blue and red represent measured and simulated spectral operating ranges and one standard deviation uncertainties of the respective mean reflectance values, indicated by horizontal lines.
d) does not show a shaded region for the experimental data as the reflectance just reaches $50\%$ at maximum values.
The orange lines show the V2 emission spectrum.
}
\label{fig:figure2}
\end{figure*}

\subsection{Fabrication and Characterisation}

The simulated reflector design was fabricated by the following recipe. The cleaned silicon carbide sample is coated with a $400\,$nm thick electron beam resist (CSAR, Allresist) and patterned by $50\,$kV electron beam lithography (Voyager, Raith Nanofabrication). After development, a nickel layer is deposited with a thickness of $150\,$ nm. The subsequent lift-off in 1-Ethyl-2-pyrrolidone (NEP) at $90^\circ$C creates a metal mask which is transferred into silicon carbide by reactive ion etching using a SF$_6$ plasma. To suspend the structures and create a triangular shape, a self-made Faraday cage with an etch angle of $\delta=54^\circ$ is used. Finally, the metal mask is removed in nitric acid followed by Piranha solution. An image of the final reflector structure is shown in Figure \ref{fig:figure2}a. On the left side of the reflector, the tapered interface is visible. 

To access the waveguide mode, the nanobeam is tapered and attached to a tapered optical fibre. With such a TWTF interface, coupling efficiencies above $90\%$ allow an efficient photon transfer into the nanostructure and vice versa \cite{Krumrein.2024,Burek.2017}.

The reflectance of the fabricated structures is measured by using the characterisation setup schematically drawn in Figure \ref{fig:figure2}b. 
It consists of a broadband light source emitting at frequencies between 286 and $428\,$THz. 
This light is sent to a 90:10 beamsplitter and then coupled into a tapered optical fibre (780HP) which is in contact to the waveguide tapers from the reflector structures.
The polarisation controller guarantees maximum coupling to the reflector structures matching the waveguide TE mode. 
The back-reflected photons are collected by the same TWTF interface and sent to the spectrometer via the second arm of the fibre beamsplitter. 
To obtain the reflectance $R$, the signal $I_{\rm sig}^{\rm R}$ from the Dinosaur reflector is divided by a reference signal $I_{\rm ref}^R$ which is obtained by replacing the tapered fibre with a fibre retroflector with a frequency-independent reflectance of $\eta_{\rm retro}=0.97$. 
The influences of waveguide transmission losses and the non-perfect waveguide-fibre coupling are corrected by a reference transmission measurement.
For this, the transmission efficiency $\frac{I_{\rm sig}^{\rm T}}{I_{\rm ref}^T}$ of a nanobeam with the same geometry is measured. 
Instead of the Dinosaur reflector, a second waveguide taper is added, which is connected by an additional tapered fibre.
With this correction term, the reflectance can be calculated as

\begin{equation}
R=\frac{I_{\rm sig}^{\rm R}}{I_{\rm ref}^R}\times\eta_{\rm retro}\times\frac{I_{\rm ref}^{\rm T}}{I_{\rm sig}^T}.
\end{equation}

The recorded reflection spectrum of the optimised Dinosaur reflector is displayed in Figure \ref{fig:figure2}c. The spectrum exhibits a $59\,$THz-wide spectral operating range from $288\,$ to $347\,$THz.
Within the operating range, the average reflectance is $R_{\rm avg}=(64.5\pm10.6)\%$ with a maximum reflectance of $83\%$ around $300\,$THz.
In the wavelength range of the V2 emission spectrum (from $286$ to $327\,$THz), the average reflectance is $R_{\rm V2,avg}=(65.8\pm14.4)\%$.
This shows that the fabricated Dinosaur reflector is perfectly suited to reflect photons emitted from the V2 colour centre.
However, waveguide-embedded colour centres have to be positioned inside interference maxima of the total electric field intensity that is the superposition of incident and reflected fields (Figure \ref{fig:figure1}a) for maximizing the fraction of colour centre fluorescence that is emitted into the waveguide mode.


To compare the recorded reflection spectrum with simulations, the parameters $A_{4,\mathrm{meas}}=164.5\,$nm and $g_{\mathrm{meas}}=63.5\,$nm are extracted from scanning electron microscopy (SEM) images of the investigated reflector.
Then, simulations are performed using these parameters and the remaining parameters of an optimised reflector as introduced above.
Figure \ref{fig:figure2}c visualises both the recorded and simulated spectra.
The spectral operating ranges of both spectra coincide closely, exhibiting similar bandwidths of $62\,$THz in simulations and $59\,$THz in the experiment.
Due to uncertainties in the parameters extracted from SEM images and possibly a small deviation in the sidewall angle from $\delta=54^\circ$, the simulated operating range's centre frequency is $6\,$THz lower than the measured operating range's centre frequency.
Within the operating ranges, the simulated mean reflectance is $19\,$\% higher than the measured mean reflectance, which may be caused by nanofabrication imperfections, such as rough surfaces or deviations from the ideal corrugation shape.
Overall, both reflection spectra show a similar shape with a small dip close to the upper ends of their operating ranges, which is induced by the second-order TE-like Bloch band separating both complete bandgaps (Figure \ref{fig:figure1}b).
Measuring reflection spectra of fabricated reflectors with different parameters $A_4$ and $g$ confirms a shift of the spectral operating range towards lower frequencies with increasing $A_4$ or $g$.

A reflector with the same number of unit cells but without a tapered waveguide-reflector interface, i.e., a reflector composed of unit cells with constant parameters $a_4$ and $A_4$ directly attached to a waveguide, possesses a worse reflection spectrum, as depicted in Figure \ref{fig:figure2}d.
Simulated as well as measured mean reflectance values drop by about $20\,$\%, leading to a $76\,$THz-wide simulated spectral operating range with a mean theoretical reflectance of $63(9)\,$\%.
Defining the measured operating range is not possible since the measured reflection spectrum barely exceeds the $50\,$\% threshold.
Despite a deteriorated performance owing to a sudden waveguide-reflector transition, the measured spectrum resembles the simulated spectrum, showing a slowly rising edge and a sharp drop at $350\,$THz, followed by a decreasing tail.
The observed reduced performance emphasises the need for a tapered interface.

\subsection{Integration of Colour Centres}

\begin{figure}
\centering
\includegraphics[width=0.475\textwidth]{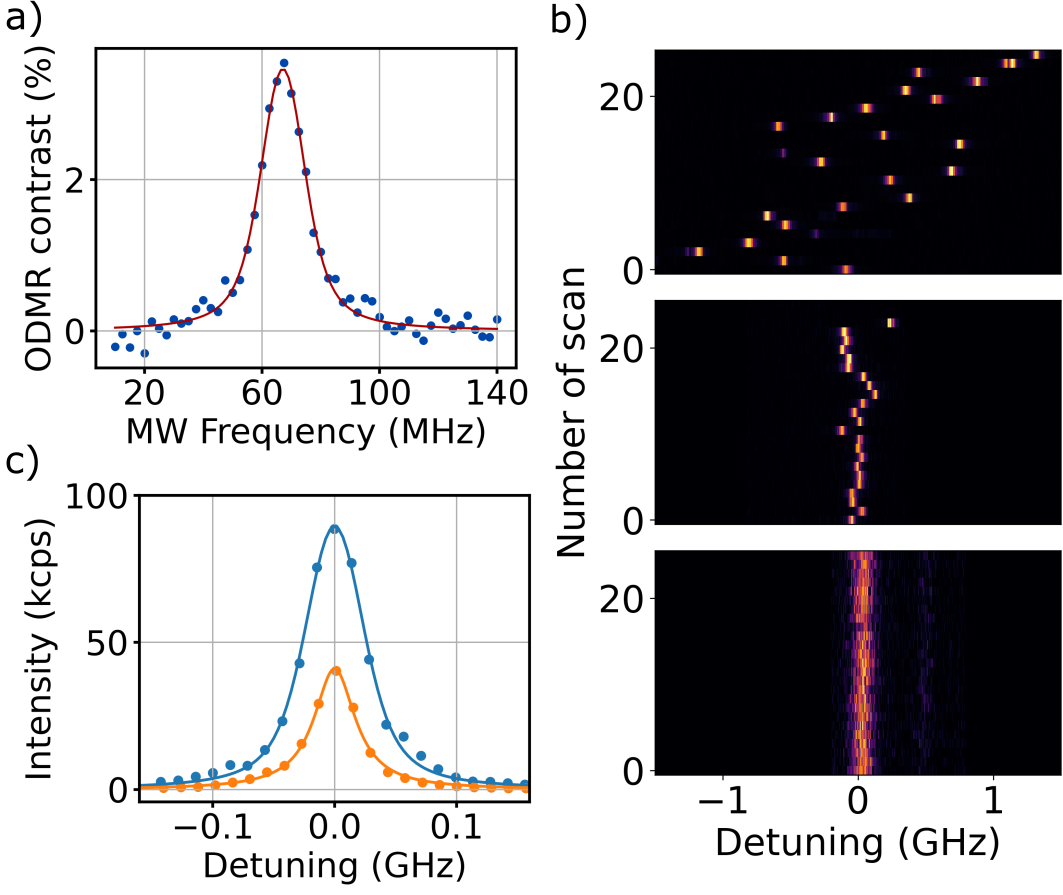}
\caption{ODMR spectrum and PLE measurements of a V2 colour centre integrated into a waveguide attached to a Dinosaur reflector at one end. 
The emitted photons are collected via a TWTF interface. 
(a) ODMR spectrum fitted with Voigt function with its peak maximum at $(67.15\pm0.19)\,\text{MHz}$ and a contrast well above $3\%$. 
(b) PLE stability over 25 scans.
Top panel: Jumping PLE lines for an emitter excited at a resonant power of $5\,$nW and a strong repump laser of $40\,\mu$W.
Middle panel: PLE scans of the same emitter but at a much lower resonant power of $0.1\,$nW and no repump applied.
Bottom panel: PLE of a bulk emitter in the  same sample as stability reference.
(b) Averaged PLE spectra of three single lines at a high excitation power of $12.5\,$nW close to saturation (blue) and at a lower power of $1.5\,$nW (orange). The fitted linewidths are $(58.2\pm1.3)\,$MHz and $(40.99\pm0.89)\,$MHz for the blue and orange curve, respectively.
}
\label{fig:figure3}
\end{figure}

Having now fabricated and characterised efficient Dinosaur reflectors, we integrate colour centres into these structures. 
For this, silicon vacancies are created in an a-plane silicon carbide sample by electron irradiation. 
Reflector structures were fabricated into bulk such that the crystal c-axis, and hence the emitter dipole axis, is perpendicular to the waveguide propagation direction. 
The photon emission was collected by a TWTF interface and the integrated colour centres were pre-characterised at room temperature. The ODMR spectrum in Figure \ref{fig:figure3}a shows the addressability of the emitter's spin states. Auto-correlation measurements and saturation studies are presented in Supplementary Note 1.

\begin{figure*}
\centering
\includegraphics[width=0.95\textwidth]{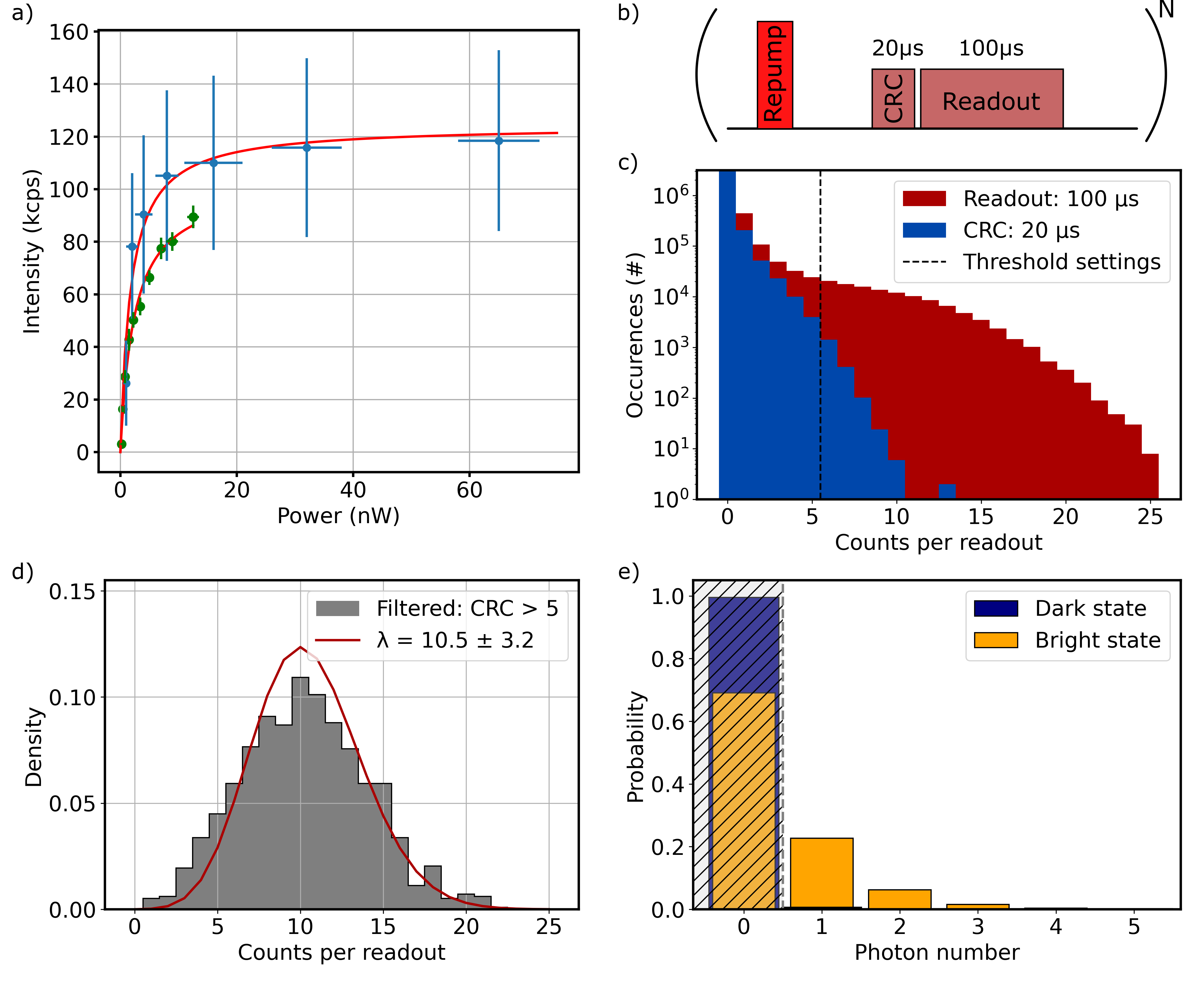}
\caption{Saturation curve and charge-resonance check measurements.
(a) Saturation curve measured by two different methods.
Blue dots: measured and rescaled counts in a $100\,\mu$s bin under the condition that the emitter was on resonance in a preceding charge resonance check.
A Poissonian distribution was fitted to the data to extract the mean value and the errorbars. The data are fitted with equation \ref{eq:saturation} revealing a saturation intensity of $I_s=(124.3\pm7.2)\,$kcps.
Green: standard PLE scans. For each power, the brightest three lines were averaged and the maximum value is taken. The data are fitted with equation \ref{eq:saturation} resulting in a saturation intensity of $I_s=(103.8\pm4.2)\,$kcps.
(b) Pulse sequence: initializing repump pulse using a $730\,$nm laser followed by a $20\,\mu$s long CRC check during which both resonant transitions A$_1$ and A$_2$ are enabled.
Subsequently, the resonant readout pulse is applied for $100\,\mu$s.
(c) Histogram of recorded events using the measurement pulse scheme shown in (b).
The dashed line marks the filtering threshold of $\text{I}_\text{CRC}>5$.
(d) Normalised histogram of the readout events conditioned on that during the CRC pulse more than 5 counts were detected. 
For the photon statistics, a Poissonian distribution was used to fit the average counts measured in the post-selected data. 
We obtain $\lambda=(10.5\pm3.2)\,$counts per $100\,\mu$s readout with an uncertainty of $\sqrt{\lambda}$, which corresponds to a count rate of $(105\pm32)\,$kcps.
(e) Histogram of a simulated optical SSR attempt with a readout fidelity of the bright state of $98.44\%$.
The calculations are based on the measured countrate for $P_\text{res}=8\,$nW.
A detailed description of the technique is given in the main text.
}
\label{fig:figure4}
\end{figure*}

\subsection{PLE Measurements}

Next, we integrated the sample in a closed-cycle cryostat to perform standard PLE measurements at cryogenic temperatures of about $10\,$K. 
For overcoming spin pumping during PLE scans, we split the exciting laser into two beam paths and energetically separate them by $1\,$GHz matching the energy difference between the A1 and A2 transitions \cite{Udvarhelyi.2020}.
We further sweep the laser frequency around $327\,$THz and collect the PSB photons (for more details, see Appendix \ref{sec:app exp setup}). 
The upper panel in Figure \ref{fig:figure3}b shows subsequent PLE scans of a single V2 colour centre excited and collected through the fibre at an elevated resonant laser power of $5\,$nW close to saturation. 
From one scan to the next one, the optical frequency of the emitter jumps several linewidths resulting in a window of more than $2\,$GHz in which the emission occurs.
This instability can be explained by reshuffling of the electronic environment of the V2 by the elevated laser power and a strong $730\,$nm repumping pulse at the beginning of each scan, which recharges the colour centre.
In particular, we expect to have many mobile charges at the nearby surfaces of the waveguide, which will be easily rearranged during the laser excitation \cite{Orphal-Kobin.2023}.
The time required for the charges to form a new equilibrium state is longer than the PLE scan itself allowing the emitter to appear either broadened or at different spectral positions.
For some colour centres, this effect can be minimised by reducing the resonant laser power and skipping the initial repumping pulse.
The middle panel of Figure \ref{fig:figure3}b shows the improved PLE stability for a resonant power of $0.1\,$nW.
Here, the spectral diffusion caused by nearby charges is noticeably less, and on short time scales of a few scans almost comparable to deep bulk V2s (see lower panel).
Waveguide emitters appear narrower compared to bulk which is attributed to less volume charges in the confined waveguide, similar to a charge-depleted bulk area \cite{Steidl.2025,Scheller.2025}.

The orange line in Figure \ref{fig:figure3}c illustrates an averaged PLE scan at a low laser power of $1.5\,$nW.
Here, we highlight that an almost lifetime-limited ($\nu_\text{lifetime}\approx 20\,$MHz) V2 colour centre with a linewidth of $(40.99\pm0.89)\,$MHz is observable in our devices.
Increasing the power leads to broader lines through power broadening, but at the same time PLE intensities close to $100\,$kcps can be achieved.
We conducted a saturation study using standard PLE measurements and obtained a saturation intensity of $I_s=(103.8\pm4.2)\,$kcps (see green dots in Figure \ref{fig:figure4}a).
This is about four to five times brighter than collected from V2s in solid immersion lenses \cite{Korber.2024}.
Presumably, these count rates are underestimated because the emitter can drift during the readout of the PLE measurement due to the fast movement of the V2's optical transition.
For high powers, the single line scans cannot be fitted properly because of increased spectral movement and ionisation processes. 
This restricts the saturation study to moderate powers.

\subsection{Charge-resonance check}
In order to avoid this fast spectral diffusion during the measurement, a pulsed readout scheme was applied (see Figure \ref{fig:figure4}b).
To guarantee that the colour centre is in the correct charge state, a repump laser pulse with an off-resonant wavelength of $730\,$nm is applied.

The laser frequency is fixed in the centre of the window of spectral wandering and PLE data are collected within a $100\,\mu$s readout window.
Those data are filtered on the initial state, thus, we apply a very fast, $20\,\mu$s long charge resonance check (CRC) right before the readout.
By that, the readout data can be post-processed and filtered afterwards conditioned on whether the CRC was dark or bright.
By increasing the threshold, more and more data are discarded. 
In return, the certainty that the emitter was bright and on resonance during the readout increases and a clear Poissonian-like distribution as expected for photon statistics is formed.
The threshold was set to 5 counts per CRC readout and the adapted data were fitted with a Poissonian function of the form $\frac{\lambda^k}{k!}\cdot e^{-\lambda}$ to extract the average number of counts ($\lambda$) and the uncertainty ($\sqrt{\lambda}$) for measurements conducted at different resonant powers.
We further extrapolate the counts from the $100\,\mu$s readout window to a count rate of $1\,$Hz and fit the data points with the saturation curve shown in Appendix \ref{supp:saturation}, leading to a saturation intensity of $I_s=(124.3\pm7.2)\,$kcps (red curve in Figure \ref{fig:figure4}a).
The overall saturation count rate increases compared to the standard PLE method (green data in Figure \ref{fig:figure4}a) because the probability of ionisation events and detuning of the colour centre is reduced.
But for higher resonant laser powers, the Poissonian fit does not match perfectly anymore, especially in cases of few counts per readout bin.
The reason is that the spectral diffusion is still faster than the readout cycle.
Hence, we expect even more counts for V2s with a spectral stability similar to bulk emitters.
Surface passivation and applying electric fields may help to reduce these drifts and further stabilise the colour centre.

As last step, we estimate the attempt of an optical single shot readout based on the obtained countrate (for $P_\text{res}=8\,$nW). 
For this, we assume that it is possible to stabilise the PLE transition without diminishing the intensity. 
In this idealised scenario, the emitter is initialised in the $m_s=\pm 3/2$ state and excited via the A$_2$ transition.
From here, the bright state $\big|b\rangle$ will decrease into the $m_s=\pm 1/2$ state via intersystem crossing with a decay rate of $\gamma$. 
If the emitter is in the dark state $\big|d\rangle$, it will remain in it because we neglect forbidden transitions in our consideration.
To estimate the result of SSR, we calculate the probability $p$ of having $k$ photons in the bright or dark state in a given time interval $T$ as

\begin{subequations}
\begin{align}
p(k\big|\big|b\rangle) &= \int_0^T \text{e}^{-\gamma t} \cdot \text{Poiss}(k, \lambda_{b} t + \lambda_{d} (T-t)) \mathrm{d}t,\label{eq:bright}\\
p(k||d\rangle) &= \text{Poiss}(k, \lambda_{d} T),\label{eq:dark}
\end{align}
\label{eq:bright and dark}
\end{subequations}

\noindent taken from \cite{Zahedian.2024}.
Here, $\lambda_i$ $(i\in\lbrace b,d\rbrace)$ denotes the mean photon rate in the corresponding state and Poiss the Poissonian distribution.
The transitions between the bright to dark states were studied in \cite{Liu.2024}, and based on their results we approximate the dynamics with a bi-exponential decay characterised by the decay rates $\gamma'$ and $\gamma''$ and corresponding weights $a'$ and $a''$. Thus, equation \eqref{eq:bright} modifies to

\begin{equation}
\begin{aligned}
p(k||b\rangle) = \int_0^T &(a' \cdot \text{e}^{-\gamma' t} + a'' \cdot \text{e}^{-\gamma'' t})\\
&\cdot \text{Poiss}(k, \lambda_{b}t + \lambda_{d}(T-t)) \mathrm{d}t.
\end{aligned}
\label{eq:bright mod1}
\end{equation}

\noindent Further, one additional readout assisted by a nuclear spin memory \cite{Hesselmeier.2024} is taken into account leading to the following modification:

\begin{equation}
p(k||b\rangle)_\text{Nuc} = \sum_{N=0}^k p(N||b\rangle) \cdot p(k-N||b\rangle).
\label{eq:bright mod2}
\end{equation}

\noindent With the equations \eqref{eq:dark} and \eqref{eq:bright mod2}, we can then simulate the SSR resulting in the histogram shown in Figure \ref{fig:figure4}e.
With a threshold of >0 counts, we obtain a readout fidelity of the bright state of $98.44\%$ (for further information, see Supplementary Note 2).
However, we have to mention that $69.1\%$ of the measurement data are discarded (shaded area in Figure \ref{fig:figure4}d). 
For other excitation powers, the simulated fidelities are summarised in Supplementary Note 2. 
Our results show that under the assumptions of spectrally stable emitters our device allows for optical single-shot readout of the electron spin using only one nuclear-spin assisted repetition, which is an important step towards reliable quantum applications.

\section{Conclusion}
We have simulated, fabricated, and measured Dinosaur PhC reflectors with a triangular cross section and corrugation features.
The fabricated reflectors exhibit a broad spectral operating range of about $60\,$THz and a maximal reflectance exceeding $80\,$\%.
Moreover, we have shown that tapered waveguide-reflector interfaces to adiabatically interconvert waveguide and Bloch modes are indispensable and need to be designed conscientiously, since such interfaces affect the performance of PhC nanostrucures tremendously.
Bandgap engineering for broadband performance is equally important as designing tapered interfaces for adiabatic mode conversion.
Furthermore, the Dinosaur reflector-waveguide structures with embedded V2 colour centres were integrated and investigated in a cryogenic environment.
Although the spectral stability is similar to bulk emitters for low excitation powers, the V2 colour centre must be further stabilised for high excitation powers.
With standard PLE measurements, we obtain a saturation intensity of $I_\text{s,LT}=(103.8\pm4.2)\,\text{kcps}$.
By applying a pulsed measurement scheme and filtering the photons according to CRC conditions, count rates of up to $125\,$kcps can be achieved at cryogenic temperatures.
Finally, we evaluate our devices in regard to successful optical SSR of the electron spin.
Under idealised conditions, optical SSR of the bright state is possible with a fidelity of $98.44\%$.
Our platform further pushes the V2 qubit system in 4H-SiC, allowing for direct SSR via the optical transition as the next step.

\acknowledgments
We acknowledge Dr. Rainer St{\"o}hr, E. Hesselmeier-H{\"u}ttmann, J. K{\"o}rber, P. Kuna and M. Hagel for fruitful discussions and experimental help.\\
J.U.H., V.V. and J.W. acknowledge support from the European Union's Horizon Europe research and innovation program through the 'SPINUS' project (No. 101135699).
J.W. acknowledges support from the German Federal Ministry of Research, Technology and Space (BMFTR) for the projects 'Spinning' (No. 13N16219) and 'QSI2V' (No. 13N16756).
J.W. further acknowledges support from Clusters4Future for the project 'QVOL' (No. 03ZU2110GB).
T.Sch. acknowledges support from the German Federal Ministry of Research, Technology and Space (BMFTR) for the 'QPIC-1' project (No. 13N15858). 
T.Sch. further acknowledges the European Research Council for the ERC Starting Grant 'QUREP' (No. 851810). 
J.U.H. acknowledges support from the Swedish Research Council (No. 2020-05444), the Knut and Alice Wallenberg Foundation (No. KAW 2018.0071), and the European Commission through the QuantERA project InQuRe (No. 731473, and 101017733).

\section*{Author Contributions}
M.K. fabricated and characterised the PhC reflector, performed the ODMR spectroscopy, the standard PLE measurements, and the saturation study.
J.B. designed and simulated the PhC reflector, and compared the experimental results with the simulations.
T.St. performed the CRC measurements and estimated the SSR attempt.
The high-quality samples were grown by J.U.H. and electron irradiated by W.K.
J.W., T.Sch. and V.V. supervised the projects.
The manuscript was written by M.K., J.B. and T.St.
All authors contributed to the manuscript.

\section*{Declarations}
The authors declare no competing interests.

\appendix
\section{Sample Properties}
We use a 4H-SiC a-plane sample which was overgrown with a $10\,\mu$m thick n-type epilayer by chemical vapor deposition.
The free carrier concentration is $4\times10^{13}\,\text{cm}^{-3}$ with an isotopically engineered abundance of
isotopes ($0.5\%$ $^{13}$C and $0.5\%$ $^{29}$Si).
Colour centres were implanted by electron irradiation with an energy of $5\,$MeV and a dose of $2\,$kGy. 
Afterwards, the sample was annealed at $600^\circ$C in argon atmosphere for 30$\,$minutes. 

\section{Saturation study}\label{supp:saturation}
For continuous-wave excitation, the emission of the V2 colour centre is limited by the metastable state. 
Hence, the saturation curve is described best by

\begin{equation}
    I(P)=\frac{I_s\cdot P}{P+P_s},
    \label{eq:saturation}
\end{equation}

\noindent with $I_s$ as the saturation intensity and $P_s$ as the saturation power.

\section{Experimental Setup}\label{sec:app exp setup}
The low-temperature measurements were performed in a  Montana Instruments cryostation. 
The colour centres are excited via a tapered optical fibre (1060XP) and the PSB emission is collected by the same fibre. 
For proper waveguide-fibre coupling, a wide-field imaging system was installed.
For initialisation and repumping, a diode laser operating at $730\,$nm (Cobolt 06-MLD from H{\"u}bner Photonics) was used.
For resonant optical excitation, we fed a mode-hopping-free wavelength-tunable laser (Toptica CTL) into the optical fibre.
We separated the resonant laser into two beam paths which we further energetically split by guiding the beams through multiple acousto-optic modulators (AOM) to induce a total frequency difference of $1\,$GHz (which is the gap between both optical transitions).
Excitation and detection are combined and separated by a fibre-based 99:1 beam splitter in a 2x2 configuration.
To filter the laser photons, a tunable long-pass filter was inserted into the detection path (Semrock TLP01-995).
The PSB photons are recorded by fibre-coupled SNSPDs from Photon Spot.

\bibliography{bibliography}


\end{document}


\title{Supplementary Information for "Waveguide-integrated colour centres in silicon carbide with broadband photonic crystal reflectors for efficient readout"}

\author{Marcel Krumrein\,\orcidlink{0009-0009-4685-8831}}
 \affiliation{3rd Institute of Physics, IQST, and Research Center SCoPE, University of Stuttgart, 70569 Stuttgart, Germany}

\author{Julian M. Bopp\,\orcidlink{0000-0003-0370-9114}}
 \affiliation{Humboldt-Universität zu Berlin, Department of Physics, 12489 Berlin, Germany}
 \affiliation{Ferdinand-Braun-Institut gGmbH, Leibniz-Institut für Höchstfrequenztechnik, 12489 Berlin, Germany}

\author{Timo Steidl\,\orcidlink{0009-0000-2819-3228}}
 \affiliation{3rd Institute of Physics, IQST, and Research Center SCoPE, University of Stuttgart, 70569 Stuttgart, Germany}
 
\author{Wolfgang Knolle}
 \affiliation{Leibniz-Institute of Surface Engineering (IOM), 04318 Leipzig, Germany}

\author{Jawad Ul-Hassan}
 \affiliation{Department of Physics, Chemistry and Biology, Linköping University, SE-581 83 Linköping, Sweden}

\author{Vadim Vorobyov\,\orcidlink{0000-0002-6784-4932}}
 \affiliation{3rd Institute of Physics, IQST, and Research Center SCoPE, University of Stuttgart, 70569 Stuttgart, Germany}
 
\author{Tim Schr\"oder}
 \affiliation{Humboldt-Universität zu Berlin, Department of Physics, 12489 Berlin, Germany}
 \affiliation{Ferdinand-Braun-Institut gGmbH, Leibniz-Institut für Höchstfrequenztechnik, 12489 Berlin, Germany}

\author{J\"org Wrachtrup\,\orcidlink{0000-0003-3328-9093}}
 \affiliation{3rd Institute of Physics, IQST, and Research Center SCoPE, University of Stuttgart, 70569 Stuttgart, Germany}
 \affiliation{Max Planck Insitute for Solid State Research, 70569 Stuttgart, Germany}


\maketitle

\section{Room-temperature characterisation of the waveguide-integrated V2 colour centres}

To characterise the integrated colour centres, the emitters are excited confocally from top by a pulsed laser diode that emits at a wavelength of $780\,$nm. 
The emitted photons are collected by a TWTF interface and detected by a superconducting nanowire single-photon detector (SNSPD). 
The entire setup is described in detail in previous works \cite{Krumrein.2024}.

V2 candidates are identified in confocal scans and verified by optically detected magnetic resonance (ODMR) spectroscopy. 
We found ODMR peaks between $67$ and $74\,$MHz, which is a deviation of maximally $4\,$MHz from the ground state zero-field splitting of $70\,$MHz \cite{Soykal.2017}. 
From these values, we can conclude that the strain components within the fabricated nanobeams are not higher than $5\cdot10^{-4}$ \cite{Krumrein.2024}. 
A representative ODMR spectrum of one V2 colour centre is shown in the main text in Figure 3a with a peak position of $(67.15\pm0.19)\,\text{MHz}$, a linewidth of $(18.23\pm0.36)\,$MHz, and an ODMR contrast above $3\%$.

\begin{figure*}
\includegraphics[width=0.95\textwidth]{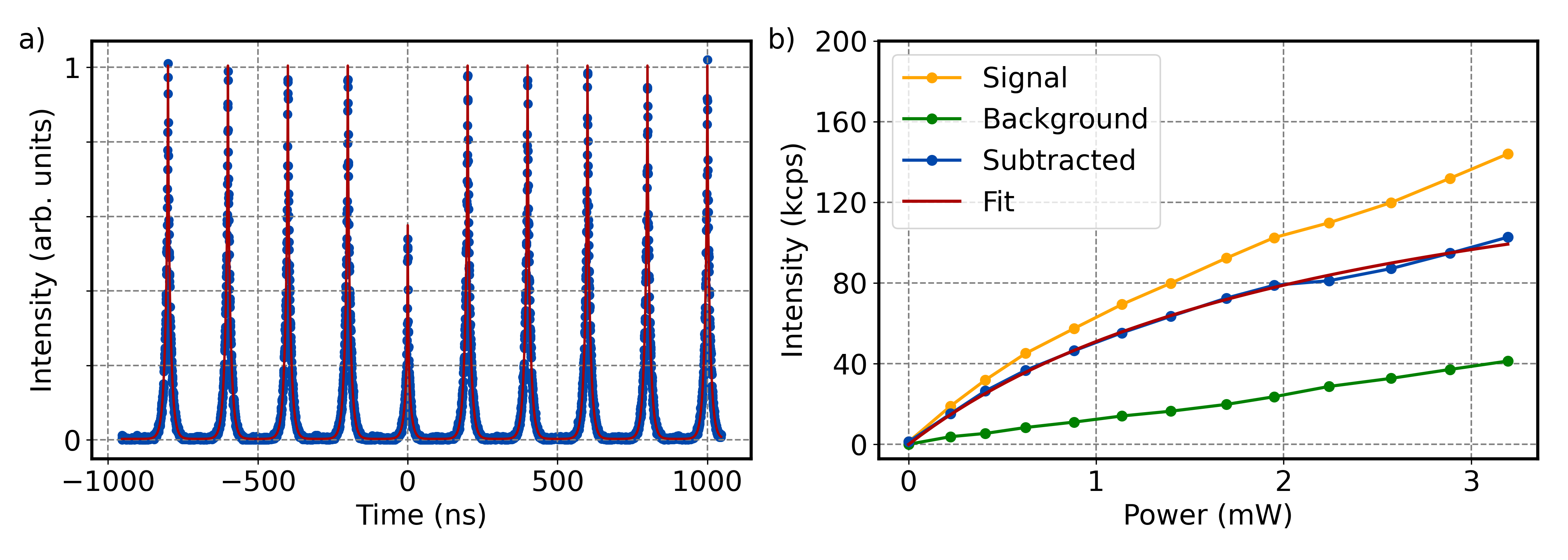}
\caption{Room temperature characterisation of a V2 colour centre integrated into a waveguide attached to a Dinosaur reflector at one end. 
The emitted photons are collected via a TWTF interface. 
(a) Second-order autocorrelation measurement under pulsed excitation. 
The peaks were fitted with a bi-exponential decay, and the area below the peaks integrated, which results in an antibunching dip of $g^2(0)=0.4100\pm0.0077$. 
(b) Power-dependent saturation curve with a fitted saturation intensity of $I_s=(174.5\pm6.9)\,\text{kcps}$.}
\label{fig:SI1}
\end{figure*}

The single-photon character of the V2 colour centre is investigated by second-order auto-correlation measurements. 
The correlation data of a V2 colour centre are displayed in Figure \ref{fig:SI1}a revealing that the peaks follow a bi-exponential decay. 
For mono-exponential decay curves, the ratio of the peak amplitudes is equivalent to the ratio of the area below the peaks. 
This is not valid for bi-exponentially decaying signals. 
To obtain $g^2(0)$, we must consider the area below the peaks. 
For this, a multi-peak fitting routing was used to obtain the red fitting curve in Figure \ref{fig:SI1}a.  
All peaks are fitted simultaneously with a bi-exponential function with the same decay constants for all peaks. 
The resulting fit matches the experimental data accurately. 
After dark count subtraction, the area below the peaks was numerically integrated.
The error for $g^2(0)$ is obtained by maximum error estimation: the fit parameters were varied within the $1\sigma$ error interval and inserted into the fit function.
We obtain $g^2(0)=0.4100\pm0.0077$ for the antibunching feature. 
Hence, the emitter can be considered as a photon source with single-photon character.

Power-dependent saturation studies were performed. 
After subtraction of the linear background, we obtain a saturation intensity of $I_\text{s,ref}=(174.5\pm6.9)\,\text{kcps}$. 
In previous works \cite{Krumrein.2024}, we showed that waveguide-integrated V2 colour centres with the same background level ($g^2(0)=0.466\pm0.003$) have a saturation intensity of $I_\text{s,wg}=(224.7\pm8.6)\,$kcps when collecting the emission at both ends of the waveguide. 
With this, we can now calculate the reflectance $R_\text{V2}$ of the Dinosaur reflector based on the saturation intensity by

\begin{equation}
I_\text{s,ref}=\frac{1}{2}\cdot I_\text{s,wg}\cdot (1 + R_\text{V2})
\label{eq:reflectance}
\end{equation}

\noindent and obtain $R_\text{V2}=(55.3\pm8.5)\%$. 
This is slightly lower than the average reflectance $R_\text{V2,avg}=65.8\%$ calculated in the reflection measurements. 
We attribute this deviation to a non-ideal positioning of the colour centre in the nanobeam's cross section.
In conclusion, the Dinosaur reflector allows us to collect $78\%$ of the photons compared to a collection from both waveguide ends.

\section{Estimation of optical single shot readout}

For the theoretical calculations of optical single-shot readout in the main text, the used parameters are shown in Table \ref{tab:supp}. Count rates for the bright and dark state are taken from the saturation measurement. To obtain the rates and weights, the data from \cite{Liu.2024} are taken and fitted with a bi-exponential decay. The power-dependent values have been mapped onto our values.

\begin{table}[h]
\caption{Parameters used for calculating the optical SSR. 
The count rates are taken from the saturation measurements. The weights (a' and a'') and ISC rates ($\gamma$' and $\gamma$'') are obtained from \cite{Liu.2024}.}
\begin{tabular}{c||c|c|c|c|c|c|c}
    parameter & $\lambda_b$ (kcps) & $\lambda_d$ (cps) & a' & a'' & $\gamma$' (1/$\mu$s) & $\gamma$'' (1/$\mu$s) & T ($\mu$s)\\\hline
    value & 105 & 490 & 0.768 & 0.232 & 1/0.48 & 1/3.15 & 10 
\end{tabular}
\label{tab:supp}
\end{table}

\begin{figure*}[h]
\includegraphics[width=0.475\textwidth]{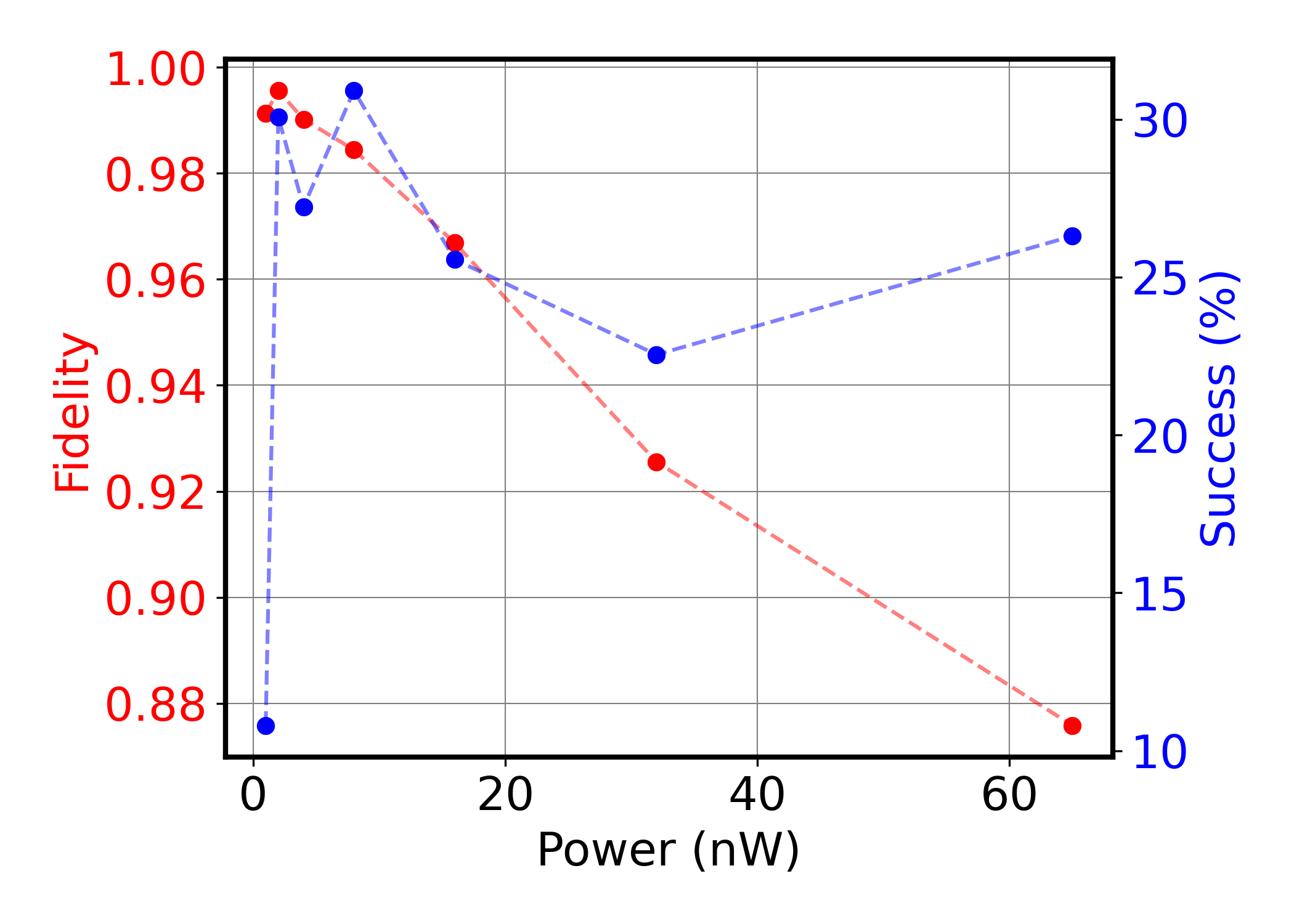}
\caption{Simulated fidelities and success rates for different excitation powers.
The calculation procedure is explained in the main text.}
\label{fig:SI2}
\end{figure*}

The calculation of the SSR fidelity was performed for all power values. 
A summary of the obtained fidelities and success rates is given in Figure \ref{fig:SI2}. 
Increasing the resonant power lowers the fidelity due to more laser-related dark counts but also does not increase the succsess rate, as the rate of switching into the dark state also rises \cite{Liu.2024}.

\bibliography{bibSI}
